# Power Dynamics in Physics

Apriel K Hodari,[1] Shayna B Krammes[1]
Chanda Prescod-Weinstein,[2] Brian D Nord,[3] Jessica N Esquivel,[3] Kétévi A Assamagan[4]

[1]Eureka Scientific Inc, Oakland, CA 94602
[2]University of New Hampshire, Durham, NH 03824
[3]Fermi National Accelerator Laboratory, Batavia, IL 60510
[4]Brookhaven National Laboratory, Upton, NY 11973

EXECUTIVE SUMMARY
The purpose of this white paper is to describe how unfair power dynamics related to various aspects of identity—race, gender identity, gender expression, sexual orientation, and ability status—operate in physics settings and offer concrete steps that one can take to make our discipline more equitable and just.

## Power Plays

Given the nature and history of racism, all white people have been socialized to hold a number of racist beliefs yet spend a considerable amount of time attempting to distance themselves from this fact. The result is racial harm done unto BIPOC people on a daily basis; their accounts of this harm are routinely dismissed because of the sincere belief that intention matters more than impact, and that "nice" people cannot be racist.

Comfortable with Oppression. Yet how nice can a person really be if they are comfortable with oppression? The evidence of oppression in all forms is abundant, yet white people remain surprised every time an account of racism or racial violence is brought to their attention. There are incentives to not knowing the true extent of oppression in the United States, and remaining comfortable, blissfully ignorant, is just one example.

Attempts to Stand Straight in a Crooked Room. Intersectionality methodology guides our understanding of the experiences of BIPOC women and allows us to see the different oppressions that they must navigate despite our society's tendency to obscure it.

## Recommendations

This paper concludes with several recommendations for ways that readers might apply the topics addressed herein. The first—to observe and learn about power dynamics and movement-building—will require you to see new dynamics with new eyes, even in contexts you previously believed were quite familiar. Equally challenging is the suggestion to examine your own values, and whether/how they align with those around you.

Finally, we offer a number of ideas how readers might use their power to disrupt oppressive structures, and re-imagine a new landscape. To make true and lasting change, both existing tools, and new tools/skills must be brought to bear. The challenge must also be met with a sense of urgency and a commitment to our future selves. Snowmass 2031 will arrive whether we make improvements or not. The question is, "How will we spend the intervening years?"

<u>A TALE OF TWO TABLES</u>

> *One day, I walked into a huge lecture hall and saw, down at the front, one of my informants, Zina, a tall, dark-skinned African American woman. She was sitting in an aisle seat; the rest of the row she sat in was empty. I sat through class with her, and at the end of class she told me that whatever row she sits in, she clears it out—no one will sit within five or six seats of her. She explained that she used to sit in the sixth row, all by herself. Recently she had moved up to the fourth row, which had previously had habitual occupants. Now, as I saw for myself when I looked around, the sixth row held a number of students and the fourth row was empty.*

> *I asked other African American students whether this happened to them. One told me an interesting story. She said that her roommate, also African American, said to her one day "let's go down to the library and clear out a table." She was puzzled, but they went together and sat down at a table in the library where several other students were working. Within a few minutes, all of them had left. From then on, my informant told me, she started to notice that whenever she sat down at a table, although no one appeared to notice her, within 15 minutes she was always the only person at the table even if all the other tables were crowded.*

<div align="right">(Johnson, 2007)</div>

Another table is in a large communal space in a physics department. It's the space where department celebrations (retirement parties, welcomes of new students, etc) are held. Most often, it's used by undergraduate students to study, eat lunch, and meet friends before/after class. It's busiest when students have problems sets due or when grad students descend to collect leftovers from a social event. It's set with long rectangular tables and white boards, and is adjacent to the department kitchen, where staff often wait for access behind groups of students heating lunch or making hot drinks.

On one of these problem-set days, a table is filled with students, perhaps 10-12 of them, almost all tall white men. At the end of the table is a small woman. She has a quiet yet clear voice. She is wearing a hijab. Students seem to be clamoring to sit at her table. The conversation at this table seems to center around physics content, and it's intense. As I walk past, I overhear the woman speaking in short declarative sentences. She seems confident. The other students seem to respect what she is saying.

*I*n this paper, we will examine power dynamics within the disciplinary context of physics. We start with a presentation on how those with privilege live comfortably alongside the oppression of others. Next, we offer and alternative analytical frame to help us see and understand how oppression operates in physics. We finish with a discussion of what we can do to create more just disciplinary and organizational environments, and some recommendations of where to begin.

<u>COMFORTABLE WITH OPPRESSION</u>

Social scientists have been documenting for decades the increasingly subtle ways people with privilege act to reinforce oppression. Current work centers on a range of behaviors and practices—implicit bias, microaggressions, white fragility, woke performativity, silencing, hyper-visibility/invisibility, gaslighting, virtue signaling—that both oppress and hide oppression, all in support of *the lie*. Here we will explore some examples of these subtleties, how they play out, how many of us (sometimes unknowingly) perpetuate them, and thereby how we support the lie, regardless of our sincere intent not to do so.

One example is the near-ubiquity of implicit bias training. Implicit bias training has been adopted by a broad range of organizations from police departments to Starbucks and other retail companies. In the retail setting, this training is usually a computer module that presents the trainee with scenarios. An example might be to select a person from a group of photos on screen for help with car trouble, and the person who is a mechanic turns out to be the woman - surprise! Not only are these scenarios silly and predictable, they do nothing to interrupt racism or white supremacy. Implicit bias training in the corporate setting tries to avoid these topics, yet many people show up to implicit bias training and think that alone is "doing the work" to address oppression.

In reality this is performative, and does little (or nothing) to change behavior, policy, or practice. For background on how implicit bias training came to be, and more on why it is so ineffective, see our accompanying white paper *Informal Socialization* (Hodari et al., 2022a). Participants in implicit bias training will likely continue to perpetuate harm, making such intervention another way to support the structural oppression it ignores while cloaking itself in the veneer of <u>anti-oppression</u>. As scholar Robin DiAngelo and somatic therapist Resmaa Menakem discussed in 2020, living in comfort in an oppressive society has a cost (DiAngelo & Menakem, 2020):

DiAngelo: *I just really want to push back against any narrative that white people are innocent of race. I think it takes energy not to see it. It's a kind of willful not-knowing or refusal to know.*

*And I offer that question. When white people ask me, what do I do? I ask them in return, how have you managed not to know, when the information's everywhere, they've been telling us forever? What does it take for us to ask, and then to keep asking? And it just speaks to how seductive the forces of comfort are.*

*As a white person—I just want everyone to take this in—as a white person, I live in a racist society in racial comfort. I'm comfortable in a racist society. Like, wow! So what am I going to do to keep myself uncomfortable, because that comfort is really seductive and powerful?*

Menakem: *And has a cost.*

DiAngelo: *Yes.*

Menakem: *It is not a seductiveness without a cost. It's that most white people are willing for other people to pay that cost.*

Tippett: *I feel like in both of your work—and this is not enough. In some ways, this isn't even an answer to the question of how to begin. But there is a necessity and virtue of white people—of me—letting myself be uncomfortable.*

DiAngelo: *But at the same time, think about the language of violence so many white people use to describe that discomfort. So we say things like, well, I'm not going to have that conversation, because I don't want to be attacked. Attacked, right? That's a—and we're only talking about some chosen* *moment of discomfort in a conversation. And what a perversion of the true direction of violence that we've been perpetrating, or in our name, for hundreds of years.*

This excerpt opens with DiAngelo referencing the phenomenon of white people knowing that they are white, yet refusing to associate any meaning or historical context to their race. To continue to live under the assumption or, rather, affirmation, that being white does not inherently mean anything is to ignore the foundations of this nation, the testimony of black folks, and the body of scholarly work; all which indicate that the opposite is true. To be white is a meaningful thing, as it grants white people the privilege of getting to pretend that racism doesn't exist, that race doesn't carry meaning, both of which keep white people comfortable, at a cost.

Menakem mentions that other people pay the cost of white comfort. A black woman pays the cost every time a white woman asks to touch her hair, or does so without consent; the white woman can walk away from that interaction without ever thinking of it again, yet the black woman will have to sit with the discomfort and violation of her bodily autonomy. To be aware of race is to feel discomfort, at least sometimes, and white people have proven to be deeply invested in avoiding that discomfort. In fact, as DiAngelo argues that at times this avoidance leads to false accusations of violence. To be called out for a racist remark is not the same as being under attack, it is simply uncomfortable in a way that white people usually do not experience. White people may sincerely believe that being questioned about racism is an attack of their moral character, but sincere belief does not equal fact. Associating discomfort with violence will continue to hinder our ability to challenge racism.

### Sincerity as a Mask for Cruelty

In the face of despair and uncertainty, our commitment to *the lie* leaves us clinging to our worst impulses. We pretend that suffering is the exception rather than the rule. And we are sincere in our beliefs, because we are so accustomed to the lie that the truth becomes too extreme for us to tolerate (Glaude Jr., 2020).

Yet sincerity is a mere mask for cruelty. How dare we critique valuing theoretical knowledge building over experimentation, just because it tends to exclude more marginalized groups? Physics subfields focused on theory building are likely more white and male than those that focus on experimentation, particularly given how white women and BIPOC people are encouraged to choose more "realistic" applied subfields (Hodari et al., 2016; Kachchaf et al., 2015). However, the cultural beliefs around these outcomes are that these are natural, personal choices, and that the existing hierarchy is based on tradition that keeps our science "pure" and gives us a clear way to ensure quality. And we are often quite sincere in our beliefs about this tradition. We center theoretical knowledge with the most prestigious awards, and the most lucrative endowed positions.

Instead, we should consider whether the practices surrounding these beliefs—what researchers call *prestige asymmetry*[1]—reify a pattern of oppression so ubiquitous it is invisible (Martin, 2017; Prescod-Weinstein, 2020). Perhaps people who are not white cisgender men just need to grow thicker skin, conform to the way things are, and deal with it, no matter how much it hurts. They will have to shape up or get out; so that our science remains pure. We will all do whatever mental gymnastics are needed to tell this story with all the sincerity in our hearts, to concentrate on the triumph of peak intellect over mediocrity. Minoritized people will continue to disappear into the background, crumpled under the weight of the lie. No human actions involved. And the lie remains firmly in place.

---

[1] More on prestige asymmetry in the *Policing and Gatekeeping* paper (Hodari et al., 2022b).

### Wrestling with Intention as Cover for the Lie

After sincerity, one of the most popular past times we engage with when confronted by a person or situation that reveal harm is to focus on the intention of the perpetrator. But by focusing on arguments over the intentions of a particular individual or group, rather than focusing on the harm caused, we ensure that the harm will continue because it is either downplayed or not addressed at all (DiAngelo, 2018; Glaude Jr., 2020).

As is discussed in greater detail in the next subsection, wrestling with intention serves as a cover, not only for the person who unintentionally caused harm—yet **caused harm** nonetheless—the distraction from the harm to the intent also **provides cover for those who DO intend harm**. Think about that. By centering white feelings (intentions) rather than harm, the harm does not get addressed, making it easier for harm to continue. Creating conditions for further harm also gives cover to those who NOT have good intentions.

### Common White Moves that Enact Racial Harm

In *Nice Racism*, anti-racist educator Robin DiAngelo describes common discursive strategies, "moves", white progressives make that support racial harm (2021). Analyses of these moves, and their role in perpetuating racial harm, reveal how they not only give cover to racist beliefs and the structures that support them, but do direct racial harm, despite the enactor's belief that they prove their position as not racist and their genuine intention to decrease harm (DiAngelo, 2021; Lee, 2021).

The vast majority of these moves rest on underlying, oft unknowing, internalized racial superiority; center white discomfort over racial harm; and strive for progressive social capital *with other white people*. These moves result in centering whiteness (comfort, emotions, lack of humility, out-woking performances); take race off the table; cause increased racial harm; and ultimately protect the lie.

DiAngelo provides many examples of credentialing, many of which involve a white person making claims about themselves that mark them as anti-racist, such as marching with Dr. King, voting for President Obama, and/or donating to an anti-racist organization. These examples are pretty straightforward and the connection between the claim and a perceived anti-racist identity is clear. Related to the practice of credentialing is making claims about living in a certain place (often New York City or similarly diverse city), which operates on the "ridiculous yet unquestioned…underlying belief that a racist cannot tolerate proximity to Black people" (DiAngelo, 2021, p. 61). The absurdity of such logic is thinly veiled.

Describing an interview that she did for a radio show in Boston, DiAngelo recalls (2021):

> [The] *host was a white progressive and represents a mainstream progressive organization -* [yet] *I was asked if I really thought that racism was an issue, given that we were in Boston. The idea that an entire city can be assumed to be "not racist" is in itself absurd. But this rhetorical question was referencing* Boston *of all places, with its history of violent riots against school desegregation and submitted in the context of a 2017 national survey commissioned by the* Boston Globe *that found that among eight major cities, Black people ranked Boston as least welcoming. Actor and comedian Michael Che described Boston as the most racist city he has ever visited. Only 1 percent* *of corporate board members in publicly traded firms in Massachusetts are Black. Black enrollment in Boston's many universities has not appreciably increased in three decades. Boston neighborhoods are among the most segregated in the US. A Black resident interviewed for the* Globe's *seven-part exposé of racism in Boston stated, "To be a black person in Boston is* [often] *to be the only one. You're aware of the racism. You're aware of the subtleties. It's like the air we breathe, if you're black." Yet being a resident of Boston was used by a white journalist as evidence of progressive anti-racist credentials, demonstrating how out-of-touch white progressives can be with historical and current realities. Pause and consider how this lack of awareness might inform our response when Black people raise race issues.*

The Boston example is an unfortunate parallel to physics. Consider the following edits to the quotation above: I was asked if I really thought that racism was an issue, given that we were working in physics. The idea that an entire discipline can be assumed to be "not racist" is in itself absurd. But this rhetorical question was referencing *physics* of all disciplines, with its history of exclusion resulting in fewer than two hundred black women in US history completing a PhD in the subject.

In physics, credentialing is claiming that there is a culture of no culture, that physics somehow elevates the humans who participate in it above racism, sexism, homophobia, transphobia, and ableism. Those who subscribe to this idea of no culture implicitly blame the greater culture (of their nation, institution, etc.) for being racist, and that those ideas cannot permeate physics. They are saying: *if* physics had a culture, *then* racism would be a problem, but since it does not, (white) physicists are exempt from any considerations of racism or doing anti-racist work, or advocating for a more diverse body of physics students and professionals. They are saying this in the face of many black physicists and physicists of color who openly disagree, who call out the unwelcoming environment in physics and the toxic culture that does indeed exist: homogeneity is a symptom of culture, and a largely homogenous group of white men will always display evidence of the greater white supremacist values of the nation at large.

As DiAngelo explains, this insistence on the culture of no culture, directly harms BIPOC people via denial and erasure of their experiences. Often, a milder version of the culture of no culture belief acknowledges the rare presence of individual racist people, but centers white anti-racist intent as the vast majority of physicists. This simply shifts the discussion from the content and impact of racial harm to a debate over whether harm was overtly intended. Ultimately, all of this distracts physicists from confronting and eliminating harm, thereby giving cover to oppressive behaviors, beliefs, and the structures that support them. Thus the culture of no culture takes racism off the table; gives cover to those who do have racist intent, at the cost of BIPOC people; and perpetuates the practices and structures that sustain oppression of all kinds.

Of course, being devoted to upholding the idea of a culture of no culture is now a tradition in physics. Another reason that white physicists are so unwilling to challenge this belief is that white people often "freak out" when asked to identify themselves by race, particularly in cross-racial settings (DiAngelo, 2021, pp. 83-84). Many scholars posit that whiteness works so well because of its invisibility, therefore "the shock of being exposed as white people triggers a kind of existential panic" (Anderson, 2016; Bonilla-Silva, 2012; DiAngelo, 2021, p. 84; Le & Matias, 2019; Reddy, 1998; Sue, 2004). It is fair to assume that most people would like to avoid existential panic. It is also fair to assume that white physicists, who are in fact white people, feel strongly adverse to admitting that they too have race.

This unwillingness of white physicists to acknowledge their race (and the role it plays in their career and success) is one of the contributing factors to epistemic injustice in physics. Epistemic injustice occurs when someone's ability to know is questioned or discredited because of a certain identity that they hold. Epistemic injustice combined with a disagreement leads to injustice-based deep disagreement (Lagewaard, 2021, p. 1580). A disagreement becomes deep when the parties involved do not agree on what qualifies as good evidence. In physics, there is disagreement about whether the environment is non inclusive, discriminatory, and otherwise problematic. The two sides of this argument, simply put, would have white, straight, and mostly male physicists arguing that physics settings are strictly about the science meanwhile people of color and LGBTQ+ people in physics argue that they have experienced some form of discrimination based on their race, gender, and/or sexual orientation. This alone is a disagreement (not deep).

What deepens this disagreement is the fact that, in the context of the US, people of color, especially women of color, are constantly sharing their experiences of racism and racial trauma only to be questioned and disbelieved by white people far too often. A recurring example of this are the viral videos of police murdering unarmed black people, often black men, and white people justifying the extrajudicial killings for a myriad of reasons. This disbelief results in white people and people of color having different criteria for what evidence actually proves that racism is a factor in a given setting. Racism is especially difficult to 'prove' also due to white fragility, and the widely-held (although subconscious) idea that to say someone holds racist beliefs is a judgement of their morality rather than an observation of the inevitable effect of socialization in the US (DiAngelo, 2018). Not only do people of different races have different ideas about what evidence is valid when trying to call out and correct racism, but the subject itself is incredibly sensitive for white people which makes productive dialogue and change very difficult.

Essentialized Indigenous People and Magical Negroes.  Another example of a common white move that enacts racial harm is the romanticization of BIPOC people, specifically black and indigenous people. This move is different from those listed above because it is not specific to white progressives; essentialized indigenous people and magical negroes appeal to the white masses, otherwise they would not be so commonly seen in films and popular culture—*Avatar, The Green Mile,* many roles played by Morgan Freeman (God, US President), and the story of the first Thanksgiving taught to almost every school-aged child. These stories were written by white writers and intended to be enjoyed by a broad [white] audience, meaning that any challenge to the racial status quo or colonialism is off limits. This is the insidious nature of romanticization. On the surface, these fictional relationships appear to be encouraging cross-racial and cross-cultural friendships and solidarity. In reality, they contribute to an incredibly intricate web of lies that uphold white supremacy.

If it seems unlikely that a racist or conservative white person would romanticize a black person, consider how a common argument in support of slavery, made *by the slaveholders,* was that the black people in bondage were better off that way and actually happy to serve the white people holding them in bondage (Fishel & Quarles, 1970). Regardless of political affiliation at the time, no one who participated in owning, stealing labor from, and torturing another human being should be considered progressive at any period in history. This idea of the happy slave is so delusional yet pervasive that it has survived into the present day, with North Dakota House Republican Terry Jones claiming in January of 2021 that black Americans are "glad their ancestors were brought here as slaves" (Linly, 2021). One cannot overstate the egregiousness of this statement. Who would be glad that their ancestors were forced to endure chattel slavery?

Jones was obviously misinformed by centuries of 'evidence' complied in order to justify slavery, then Jim Crow, now mass incarceration. No white person who grows up in the US can completely avoid all of this 'evidence' but, clearly, Jones took much of it to heart and without questioning anything. He is comfortable with oppression and therefore has no reason to ask questions or challenge the status quo.

In 2017, while the protests against the Dakota Access Pipeline (DAPL) at Standing Rock Indian Reservation were ongoing, Jones showed his commitment to keeping whites comfortable. Jones told *The Bismarck Tribune* in February 2017 that "'the Judiciary Committee is working really hard to balance the rights of North Dakota citizens to protest and the rights of North Dakota citizens to live under rule of law and conduct their day-to-day activities'" (Forum News Service, 2017). Despite his diplomatic wording, what Jones is really saying is that he will refuse to stand up against the oil industry to protect the water supply of the Standing Rock Sioux people. Most of the anti-protest bills were passed by North Dakota's Legislative Assembly House (Forum News Service, 2017). These lawmakers abandoned the romantic idea of essentialized indigenous people the moment that the residents of Standing Rock decided to exercise their right to protest sacrificing clean water for profit, a decision which they did not make for themselves. One who is romanticized can so quickly become villainized in this situation. In response to DAPL protestors wearing face masks, a bill was introduced to criminalize mask-wearing. Rep. Al Carlson went as far as to compare indigenous water protectors to [enemies] at war with the United States, saying "'[wearing masks] might be legal in Baghdad but not in Bismarck.'" According to Carlson, indigenous people exercising their first amendment rights should be treated like hostile combatants (Smith, 2017).

As mentioned, romanticization is not unique to any particular faction of white people; however, different groups have different ways of engaging. DiAngelo discusses the self-proclaimed "spiritual, not religious" people in *Nice Racism* (2021, pp. 111-119). This group can be found travelling to Latin America in seek of indigenous wisdom or healing. They often construct transactional relationships with certain individuals in exchange for access to that individual's traditional rituals, ceremonies, land, etc. They then use the common white move of credentialing, citing that their proximity to indigenous peoples exempts them from being racist against any group, without being able to name the tribe that hosted them. They will also claim that their spiritual knowledge or awareness has elevated them to a level of being above race, racism, and prejudice, without being able to name whose tribal lands they occupy at home in the US. They turn indigenous peoples into a single, monolithic group while they appropriate practices from many different and distinct tribes, then use elements of these practices in profitable spiritual seminars lead by and for white people, who believe that their spirituality is or should be universal (yet another common white move to enact racial harm).

While addressing the romanticization of African Americans, DiAngelo explains how romanticization can exist alongside virulent anti-blackness. The deeply ingrained ideas that most white US Americans have about Africa are that it is one monolithic place rather than an entire continent composed of different countries and cultures; Africans are essentialized similar to how Indigenous peoples in the Americas are essentialized. This means that a certain degree of spirituality or mysticism can be associated with African Americans, which has become a popular trope for filmmakers to explore. In 2001, Spike Lee coined the term "magical negro" to describe the trope of the magical black character who saves the white protagonist using supernatural skills and with fierce loyalty. Lee is very critical of this character, as well as the actors who agree to portray them. Following the release of *The Legend of Bagger Vance* in 2000, a film starring Will Smith and Matt Damon, Lee was appalled. He noted that in Depression-era Georgia "Black men were being castrated and lynched left and right. With all that going on, why are you fucking trying to teach Matt Damon a golf swing?" (ABC News, 2006; Barnes, 2011; DiAngelo, 2021, p. 118) Lee's criticism leads one to consider if the magical negro trope functions as a message to African Americans, that they should be apolitical and grateful to live in the so-called greatest country in the world. While less abrasive than Representative Jones' statement, for example, the magical negro trope constantly reinforces the idea that African Americans have a particular place in society, one of servitude, and that this should be an unquestioned source of contentment - yet another manifestation of *the lie*.

<u>Making More Efficient Racists</u>.  This subsection articulates the risks of superficial and tenuous understandings of oppression, combined with the urgency to "fix" racism, resulting in expectations of quick solutions and the drive for "best practices" to get us to the post-racial finish line quickly.  Rather than achieving any of these misguided goals, this approach amount to box-checking and superficial language incorporation, sidestepping any real change.  There is no quick and easy path here.  It is hard and messy, lest we simply make more efficient oppressors.

$M$any institutions and companies desire to be perceived as liberal, especially after the international outrage sparked by the murder of George Floyd in May 2020. Before continuing, it must be clear that a liberal *perception* was more important than actually creating an environment in which BIPOC people could thrive and racism, homophobia, sexism, etc. would not be tolerated. Any institution or company that wanted to create such a workplace would have likely started their efforts much earlier.

The implementation of race-neutral policies is often thought of as a quick and easy solution to racial bias in the workplace. Race-neutral policies include measures such as drug testing, background checks, standardized tests, and dress codes. Each of these policies, however, are products of—or can be linked to—systemic racism.  Implementing such policies without taking this into consideration will do little to help BIPOC people enter and remain in a given space.

For example, DiAngelo notes that 9.2% of African Americans have reported being reprimanded or fired for failing a drug test, compared to only 4.4% of whites. Drug-related criminal charges are also more often pursued and when the person in possession of them is black or Latinx; these groups are also convicted at higher rates for the same crimes compared to their white counterparts (DiAngelo, 2021, p. 160; Posner, 2018). Despite evidence that "Hispanic and [w]hite students were more likely to report drug use and abuse than Asian and African American students prior to coming to college and during college", these discrepant impacts on employment and law enforcement will then affect who can pass background checks to gain access to certain job opportunities and promotions (McCabe et al., 2007).

Dress codes have long been accused of having rules which disproportionately control how women and girls dress, starting as early as grade school. Of course, the rules are more likely to be enforced against black women and girls, especially those whose bodies are curvy. There is already a problem with the way black children are perceived by whites as older and therefore more threatening and less entitled to make the same mistakes as white children. For black women, their own bodies are weaponized against them as they are accused of being "too sexy" for a professional setting, and blamed for wearing clothes that do not completely hide their bodies (Joseph, 2016; Nelson, 2018).

Another common move that organizations started making in 2020 was creating DEI committees. It suddenly seemed like every organization had a brand new DEI initiative and superficial plan to make themselves look 'woke.' Random employees of color across these organizations were called upon to hold DEI seminars for their coworkers, often alongside a (white) manager or supervisor of some sort. There are many issues to unpack here, the first being that this model places the responsibility of racialized people to bear the responsibility of doing race work; it enforces the idea that racism is a problem for them to figure out for themselves, while whites focus on other things. However, many white progressives are interested in showing interest in DEI work, volunteering to help with their organization's efforts and then using this to give themselves cover when they inevitably are called out for problematic, racist behavior. Again, this is not an assessment of character but an inevitable part of socialization in the US.

Growing up white in the US also means that thinking about race or racism is not required of you, and the vast majority of whites make it quite far in their careers before anyone confronts them about it. Doing this intervention is part of DiAngelo's job, and she finds that one of the most frequently asked questions following her presentations is "what if a person of color is wrong and it's not really racism?" (2021, p. 163) This scenario is so improbable given the history of racism and white supremacy in the US, which affect everything from surviving birth (and surviving giving birth) to when we will die and from what cause. Race affects everything, but the information that white people receive starting from early childhood is that the opposite is true.

This may explain why one would ask "what if" prior to attending DiAngelo's presentation or hearing the testimony of a black coworker or student. The question becomes much more insulting after. It questions the lived experiences of people of color and the decades worth of research which supports these lived experiences. To ask "what if" is to engage in epistemic injustice and move towards injustice-based deep disagreement, because you are questioning the validity of the evidence in addition to the truth of the claim itself. To ask such a question publicly may cause further harm, should this lead to the reputation of the person of color being negatively affected.

Opening up about experiences of racism in a predominantly white environment leaves the testifier vulnerable to more than just a tarnished reputation. Speaking up is often emotionally and physically taxing, and sometimes even dangerous (DiAngelo, 2021; Eddo-Lodge, 2017; Oluo, 2019). What does the white listener lose by accepting this testimony as the truth? Even on the incredibly rare occasion that the person of color is perceiving racism to be somewhere where it is not, DiAngelo argues that the only thing that a white person will have at stake is their pride, and that to defer is simply the right choice to make (2021, p. 165).

Although this section is presented as a collection of behaviors and narratives that are (oft) unknown to those who enact them, many researchers and cultural critics notice that the enactors seem to work very hard *not to know* things that challenge their worldview (*Race and Epistemologies of Ignorance*, 2007). As DiAngelo articulated (DiAngelo & Menakem, 2020):

> Tippett: *There's someplace that I've heard you observe that—and this is a question I've asked, and it's a question I hear asked a lot—white people saying, which just confirms what you're saying, how did I not see this? And you've said we don't see it, and we do see it, but we can't admit we see it, and that this creates an irrationality.*

> DiAngelo: *We're so invested in not seeing this, for so many reasons. Yeah, it's this really weird—I'm going to imagine—you tell me, Krista, if you can relate. On the one hand, we really don't know. We really are just oblivious to this, and we're shocked when we finally see it. And, on the other hand, yeah, we know. We know. I know. You know. Both those things are actually, simultaneously true or real. And then you add that you can't admit that you know, and it makes us fairly irrational. You can add a lot of other things, too, like internalized superiority that we can't admit to, and etc.*

The point of all of this is to keep white people comfortable with oppression, and thereby ensure it continues, even when their expressed intention opposes it. This is the insidiousness of systemic and pervasive oppression. It is baked in. So the next question arises. If you have lived your life in comfort with oppression, how do you learn to see reality? In the next section, we offer an approach.

## ATTEMPTS TO STAND STRAIGHT IN A CROOKED ROOM

In her exploration of black women's political and emotional responses to common stereotypes, Melissa Harris-Perry analogies her subjects' efforts as attempts to stand upright in a crooked room, similar to how cognitive psychologists describe people's attempts to do so in physically tilted spaces (Harris-Perry, 2011; Witkin et al., 1977). While she does not address STEM directly, Harris-Perry's comparison is apt. She offers rich images of women's struggles to live as full (upright) versions of themselves in a culture that reduces them to myth.

In order to see oppression in a culture that structurally obscures it, and keeps us comfortable while doing it, we can apply the alternate analytical frame of intersectionality. In effect, we shift our gaze by centering those who experience oppression in multiple forms. This allows us move past sincerity, intention, and discursive strategies commonly used to distract us from the harmful impacts of oppression. We take off our rose-colored glasses, to see the world as it actually is.

In the thirty years since intersectionality was coined, scholars have primarily used it as a theoretical construct to describe the lived experiences of people who hold multiple disadvantaged identities. In this section, we use intersectionality as an analytical methodology, to help us make sense of how oppression is enacted, and to point to opportunities (addressed in the next section) for us to create more just systems instead.

### Intersectionality Methodology

Chayla Haynes and colleagues summarize Kimberlé Williams Crenshaw's three-dimensional intersectionality framework as "[supporting] researchers, activists, and policymakers in the complex examination of the micro and macro power dynamics shaping" the lived experiences of people with multiple marginalized identities (2020). Crenshaw's model addresses the impact of power at the structural, political, and representational levels, with specific focus on how racism, sexism, and classism co-create complex latent power dynamics in higher education (Cho et al., 2013; Crenshaw, 1989).

Similarly, Patricia Hill Collins' Domains of Power model articulates the interpersonal, cultural, disciplinary, and structural aspects of power, wherein individuals, values, rules/policy, and laws, respectively, reify belief (Collins, 2015). When applied to STEM spaces, these domains help researchers focus on how the same setting is experienced by different people, depending on their identities relative to those with the most power (Johnson, 2020). In this way, an intersectionality framework is used to investigate features "*of the* [STEM] *setting*" (emphasis in the original), rather than internal individual experience. Both models, grounded in critical theories and analytical methodologies, center on women of color. However, the methodology can be used to examine disciplinary settings, thereby revealing the impact of broader systems of oppression.

As described in detail in the accompanying white paper *Policing and Gatekeeping*, people of color cannot simply leave their racial identity behind when they enter a physics setting (Hodari et al., 2022b). The issues that plague the US as a nation are the same issues that make physics a non inclusive field; the greatness of the science cannot overcome the severity of racism and anti-blackness on its own. Racism in the US is systemic and exists within every power domain: interpersonal, cultural, disciplinary, and structural (Collins, 2015). Physics is not exempt. For this reason, the lack of inclusion in physics, more specifically the unwillingness to become an inclusive environment despite many negative stories and data, means that we are dealing with injustice-based deep disagreement. Not only do white physicists and physicists of color disagree on what counts as evidence of a negative environment, but this disagreement is based on testimonial injustice, or denying the credibility of the testifiers because of their oppressed identities.

Dr. Nadia E. Brown is a member of the #MeTooPoliSci collective and a principal investigator for an NSF ADVANCE project by the same name. In 2019 she published a commentary/article about her experience with sexual harassment at her first tenure-track position. Her harassers specifically said that she was targeted because of her race; they were counting on the fact that she would not turn against a fellow black person, especially not a black man with more power (Brown, 2019, p. 166). The harassment also took place exclusively under the guise of wanting to provide mentoring opportunities for junior faculty such as Brown. These men were well-known within their department and Brown, like many women of color, feared that no one would even believe her if she came forward. In fact, Brown knew that the African American studies director, also a black woman, consistently blamed women for the sexual harassment they endured (Brown, 2019, p. 167).

This combination of factors left Brown feeling alone, powerless, and depressed. At the time, her institution "did not have an ombudsman nor a formal reporting mechanism for faculty" and its Title IX office was dealing with the fallout from mishandling a student's claims of sexual assault (Brown, 2019, p. 167). Brown's physical health suffered as much as her mental health, as she lost an alarming amount of weight and hair over her first year at this institution (2019, p. 168). It wasn't until she found the right person who was willing to listen to her and *believe* her that things started to improve, and she was eventually able to relocate to a different institution with the tools she needed to navigate informal networks and seemingly inevitable future challenges of junior, black woman faculty.

In Brown's story, we can see three of Collins' Domains of Power at work: interpersonal, cultural, and structural. Brown witnesses the close friendships between the director of African American studies and the men harassing her, which is one of two reasons that the director is unsafe to report to. In discussions with the director, Brown also learns that she engages in victim-blaming, a by-product of cultural ideas about women being inherently sexual and provocative towards men. What victim-blaming fails to recognize is that sexual harassment is always about power rather than sexual attraction. In fact, Brown's harassers explicitly targeted her because of race not because of their own preferences but because they knew that they could not get away with harassing white women, whose accounts are historically believed over those of black men especially when it comes to sexual harassment and sexual assault. Brown's harassers understood what would be permissible in the cultural domain, and were also aware that a black woman's testimony would not be believed over that of a black man (if she even decided to report).

Interpersonal and cultural domains were also helpful to Brown in escaping harassment. She credits her dissertation co-advisor, Jane Junn, with not only listening to her but to helping her find a community of women who would listen, too (Brown, 2019, p. 170). After confiding in Junn, Brown found herself working within a positive cultural domain: an informal network that advises its members about unsafe people to avoid. Not all women are able to make the same recovery as Brown; even she recognizes that networks such as the one she found are informal and can unintentionally leave out vulnerable people (Brown, 2019, p. 172). However, formalizing something like this could be quite threatening to predatory, powerful men and is perhaps not yet possible. A network like this where the stories of marginalized women are believed and acted upon also challenges all of the ideas that have led to the creation and continuation of testimonial injustice. In other words, this informal network is not only a threat to predatory individuals; it threatens to dismantle the entire structure that enables predatory behavior in the first place.

Finally, the structural domain working against Brown in this story is the obviously broken Title IX office. There was an unacceptable lack of support for Brown to pursue within her institution, forcing her to try her luck and search for her own help. As discussed in greater detail in the accompanying white paper on *Informal Socialization*, whether or not a person finds an adequate support network is often a matter of chance (Hodari et al., 2022a). When happenstance doesn't happen, bright minds often leave academia in search of something more sustainable for their mental health.

In 2011, Angela Johnson, Jaweer Brown, Heidi Carlone, and Azita Cuevas published "Authoring identity amidst the treacherous terrain of science" which profiles three women of color (black, Latina, and American Indian, respectively) who have found success in their respective STEM fields. The article focuses on how these women authored a science identity, persisting through the challenges that they faced due to many factors of their identity such as race, gender, economic class, and religion. While Collins' work on the domains of power was not yet published in 2011, Johnson describes the process of ascribing identity onto an individual as "personal, institutional, or societal" (Johnson et al., 2011, p. 9). This demonstrates an understanding of the different levels of power at play in the lives of these three women that correspond to Collins' interpersonal, disciplinary, and structural domains.

Johnson describes science identity formation as an ongoing process of making a bid, a declaration of how one views themself, and notes that this bid must be recognized and accepted for that particular identity to stick (2011). A given person interested in science will often have to start this process of identity formation in high school or before. All of the previously mentioned factors, especially race and economic class, will affect if a high school student can even pursue a four-year degree after graduation, as well as what kind of advice they receive about which field to study and so on. At any stage in their education or career, "those whose bids for recognition consistently fail will likely abandon attempts to author those particular identities" (Johnson et al., 2011, p. 10). Whether you are a student like Johnson's informant, Alethia, who was persuaded by an interviewer's coded messages about what is required of a civil engineer or a junior faculty like Nadia Brown who finds herself in an unsupportive and, frankly, dangerous institution, experiencing a location or field that does not allow you to grow into your full potential will almost always push you to a new location or field that does. Both Dr. Brown and Alethia find themselves changing on an individual level in order to succeed. This should not be necessary. It is possible to create a setting where women of color thrive in science, even in physics specifically.

In 2020, Angela Johnson published a chapter in *Physics Identity and Gender* that examines a physics setting where women of color students and women faculty are successful. This work is similar to "Authoring identity" as it focuses on finding success in a field that presents substantial challenges, both intellectually and culturally. What distinguishes the book chapter, though, is its focus on the *setting* rather than the individual marginalized students. In fact, the chapter is structured entirely around Patricia Hill Collins' Domains of Power model, analyzing the physics department from each of the four domains using interviews with students and faculty as well as observations from Johnson's time immersed in the department.

What she found was that, in every domain, physics students valued collaboration over competition. On the interpersonal level, a male student who did not enjoy group work acknowledged that it was indeed beneficial to him, and the majority of the women interviewed shared that they enjoyed group work both in class and outside of class. The faculty shared this enthusiasm for group work and actively encouraged it by using research-based, interactive teaching methods, and so on. Given the small number of women in physics, particularly women of color, Johnson acknowledges that her findings are not very intersectional as the small sample size does not allow for the experiences of women of different races to be adequately compared. However, what seems to be the case time and time again is that what is good for women of color students is good for all students (Maton et al., 2009; Sto Domingo et al., 2019). This is to say that the traditional view of physicists as lone geniuses who do everything on their own is not the best method to follow to create more physicists, even for white men (Garland, 1993; Hrabowski III, 2015).

PUT YOUR BIG KID PANTS ON

Our research environments are embedded in a society that is in part founded in white supremacy, which fundamentally affects the power dynamics in those spaces. We are all agents with roles of supporting or changing these oppressive dynamics. Therefore, it's critical for us all to study the power structures in our workspaces, including white supremacy. In this section, we discuss stories, lessons, and ideas for how to wield power to drive change. We discuss the shape of the power landscape in academic communities, pitfalls in how people often conceive of it, and then actions that people can take to use their power to materially improve the lives of black people in STEM.

### *How Power Operates in Physics Settings*

From some vantage points, it may appear that career progression coincides with and causes a monotonic increase in power within the academic system. It can appear as if one eventually reaches a point where they're free of constraints, or that those who are starting their careers have no power at all. In reality, the landscape of power is complex and varied in academia. It evolves as one changes roles in their communities: there is generally an increase in power, but the form of it also vastly changes. But one never removes themselves from this landscape. Moreover, the nuances, complexities, and impacts of the power dynamics are typically greater for people from marginalized communities, and especially those with intersectionalized identities: they never truly escape racism and misogyny. When President Obama ran for office, Prof. Michelle Obama feared for his life: arguably the most powerful head of state in the world was still subject to violent racist threats on his life (Cobb, 2014). After Lebron James rose to the top of the NBA, he still had racists epithets painted on his home (McKirdy, 2017). Power resides in our organizational structures through policy, which also defines the roles of individuals and groups. People can also act in ways that challenge these dynamics encoded by those structures. Deeply understanding how these power dynamics intersect with or emerge from racism and misogyny is essential if one is to play a role in addressing those issues.

Most spaces in the power landscape in academic STEM (and thus physics) are well described by hierarchy, while a few can be described by collectivism. Hierarchies present many opportunities for individual actors to exercise power to great effect: individuals are bottlenecks or sieves of information flow and action. Departments, and more generally academia, is designed so that individual professors (principal investigators) are leaders in their field, and the need to distinguish oneself contributes to tendencies toward local isolation. Together, these factors encourage and often demand individuated or isolated action – not just in technical research, but also in the governance and management of communities. In physics, there is also the generally held understanding that physics is at the top of the STEM food chain, which would place physicists at the top of such a hierarchy, seemingly affording them a grander purview.

Collectivism, on the other hand, is based on the capacity of a group to act together and prioritize its goals over the goals of the individual, whether through consensus- or consent-driven decision-making. Power resides in the group as a whole, rather than in individual actors. It most often appears within particular levels of the hierarchy as based on job status: student, postdoc, faculty, etc. There is a marked preponderance of collective action at the graduate student career stage. We will next discuss these power and organization paradigms, including examples of their effects and how they relate to white supremacy in physics research spaces.

### *How to Stop Running with ~~Scissors~~ Power*

<u>Learn</u>. First, professors and principal investigators manage groups of people, and progressing through STEM careers is strongly correlated with network/community access and reputation, which go through professors and principal investigators (PIs). Moreover, these community networks are small and densely connected. Due to this, an individual PI has significant power over the other individuals in their group by controlling the access to communities and how an individual's reputation is presented in those networks. Bullying itself is known as career progression tool (Täuber & Mahmoudi, 2022):

*An emerging body of research suggests that mediocre academics in particular resort to bullying, to remove their competition.… Members of underrepresented groups report they are the targets of bullying with the intent to sabotage their careers. Some anecdotes suggest that bullies spring into action when their targets become too successful for their liking—and thus viable competition—[or you might say, when the pet becomes a threat].*

In addition to mentorship/advising relationships, this manifests within departmental decision-making, wherein there is often the tantamount of veto power.  For example, individual faculty members can scuttle hires, tenure cases, or new policies within a department; departments are often consensus-driven.

Observe.  Next, we examine the effects of these structures and dynamics, especially with respect to racial justice. The hurriedness with which many physics communities have approached issues of justice and equity reflects a misunderstanding of how to generate long-term social change. That is, white physicists tend to make decisions about when an issue related to racism is urgent and how much time the community should dedicate to it.  For example, the fast and widespread adoption of the White Supremacy checklist (Okun, 2021) reflects a misunderstanding of the problem that we face – its long history, its deep ingraining in our communities. It also seems to reflect a supposition it can be encapsulated enough that it could be addressed in one go with a checklist. The speed of the adoption relative to the willingness to read primary sources or do other deeper investigations may also reflect that physicists think that power resides in objective truth and not in one's decisions about what they do with it (Lee, 2021). This kind of misunderstanding of their power over time and misprioritization of their own sense of urgency allows white physicists, who hold a significant amount of power in the community, to be bottlenecks to real, long-lasting change.

Similarly, the generation or adoption of new programs without proper research and preparation reflects a sense of existing beyond the rules, knowing everything there is to know already. Rushing to perform only short-term actions on the timeline of or with the momentary urgency of the majority group reflects misunderstanding of how power needs to be applied; power and energy needs to be applied systematically and at the appropriate junctures, but there isn't necessarily any prescribed end to that work.  It may be indeed that we approach freedom merely asymptotically. In Angela Davis' 2013 speech "Closures and Continuities" at Birbeck University, she calls on us to know and understand the multi-century effort in the U.S. to achieve freedom for black people, to recognize the "temporal continuities" from movement to movement: "freedom is a constant struggle" (Davis, 2013; *Song of the Freedom Singers*, 1964).

From a perspective of leadership and change management, Brene Brown notes that "you can't skip day two." The first part of a long effort (like the first part of a workshop) is about getting set up, sense-making, getting an initial understanding of the challenge you're facing. The next part of an effort (or day two of a workshop) is about coming back and getting into the heart of the work (Brown, 2017; Brown, 2020).

These structures also tend to sequester information with individuals. Knowing how a system such as career progression works or who to talk to about internal policies are examples of information that are  critical for navigating or changing the system. Information asymmetry is a signature of an uneven power dynamic. In communities where information is not well archived and curated, the sequestering of information exacerbates this asymmetry. Black physicists experience extreme isolation within primarily white institutions, where there also exists a lack of appropriate mentoring and advising. Additionally, black physics communities are segregated from the majority. These work in tandem to promote information asymmetry and a power dynamic that disadvantages black physicists in the pursuit of professional STEM careers. It is an exercise of power for individuals and groups to have information and the choice of whether to share this information—including the inactive choice to not share.

It is not uncommon for individual actors in STEM departments to reserve their social or policy-related action for when they progress into positions of relative power in the hierarchy, like department chair or school dean. However, these roles remain limited, although the constraints upon these roles differ from the constraints applied to people lower in the hierarchy. For example, a department chair may be constrained by the policies of levels further up the hierarchy or by the requests of donors and benefactors. Moreover, for scientists from marginalized groups, the additional constraints of racism and misogyny exacerbate the power differential. Hierarchies don't benefit black people in STEM the way people expect, because they don't automatically progress in the same ways their counterparts do. Dr. Clifford Johnson has been critically underfunded throughout his career, never been awarded enough funding to even hire a postdoc (Feder, 2020). Brian Nord has experienced racist harassment from white postdocs (during official meetings) and grad students (on social media). We don't leave our humanity at the door of the institutions or as we climb ladders of supposed success. Without an accurate understanding of these power dynamics, academics that seek to save their change-aimed activities for when they achieve these roles may be surprised at the power they lack relative to their expectations. This is a critical part of reality that many in the academy have a limited understanding of. Furthermore, climbing ladders changes the power landscape, but one does not ever achieve power without consequence.

<u>Align</u>. In "We need to talk about systemic change," Angela Davis observes that Obama's election was naively considered to herald a post-racial America, noting that "of course it didn't make a great deal of sense that the election of one person could transform the impact of racism on institutions and attitudes of an entire country" (Davis, 2015). Moreover, Davis reminds us that "it's often those structural elements that aren't taken into consideration when there is discussion about ending racism or challenging racism." So, how can we imagine that individual actors can be the prime or final movers when it comes to ending racism in physics? How can a hierarchy of isolated individuals reckon with these rigid structures? Davis again: "... we have to talk about systemic change. We can't be content with the individual actions" (Davis, 2015).

Movements provide an alternative path generating lasting systemic change, as born out by the U.S. Civil Rights movements of the 1960's and 1970's (Davis, 2015). These movements were collective efforts that tended to emphasize consensus- or consent-based decision-making across the organization. Modern grassroots student organizations, including graduate student unions, often use this kind of structure. Similarly, groups like Particles For Justice, Change Now, University of Chicago's Astronomy and Astrophysics Inclusion, Diversity and Equity in Astronomy (IDEA) group, the Physics and Astronomy Anti-Racism Coalition (PAARC), and Not Again SU function as collectives focusing on social justice in academic and research spaces (#notagainsu, 2022; Berry et al., 2020; IDEA, 2018; Nord et al., 2020; PAARC, 2020). These groups brought together individuals from diverse cultural backgrounds and varying levels of career status from multiple institutions to strategize for new modes of information flow and change with respect to racial justice in academic spaces. Davis also reminds us that coalitions amongst people who are seeking freedom from oppression are critical for building movements that can reach critical mass (Davis, 2015):

> *Well I think we constantly have to make connections … I think we have to engage in an exercise of intersectionality. Of always foregrounding those connections so that people remember that nothing happens in isolation. Collectivism through coalition-building is an underutilized mode of organization and power exercise in academic spaces: incidents are not isolated, and they are not all obvious, so our response must not be isolated, and the exercise of our power must not be isolated.*

Finally, inaction is an exercise of power and status. If you're not affected by these issues and therefore do not feel compelled to address them, then you are requiring those who don't have your privilege to do the work. So what can you do?

- First observe and learn about power dynamics and movement-building in many communities, including yours. Study systems of power and mechanisms of oppression and marginalization, like intersectionality (Collins, 2015; Crenshaw, 1989, 1991). The people at the top don't have all the power; the people at the bottom are not powerless. The key is figuring out which kinds of power reside where, because power in many cases is more about a web than a line.

- Second, examine your values and how they align with those around you. This will permit you to build coalitions and develop strategies for exercising your power – either as a group or as individuals.

- Third, act: use your power in ways that disrupt the oppressive structures in our systems and help re-imagine a new landscape.

<u>Using Existing Tools</u>. There are several existing tools that can help us address injustices in STEM organizations, and many of them are underutilized. For example, the Amerian Association for the Advancement of Science (AAAS) has a program on Diversity and the Law (AAAS, 2022). It presents resources to help organizations enact policies and practices aimed at increasing diversity within the constraints of recent US Supreme court rulings. Among these resources is advice on under-utilized federal and state law that can be used to support diverse faculty hiring, particularly when institutions, localities, and states have histories of race/ethnicity and/or gender/sex (RES) discrimination. Included in these are protections against de facto discrimination, based in part of persistent organizational or disciplinary under-representation.

Another tool is publicly-available data on representation in educational attainment, such as those data collected by the Integrated Postsecondary Education Data System (IPEDS), which can be used to make the case for action (NCES, 2022). An example of how these can be applied is the portal under development by researchers to present inclusion for women of color in physics, mathematics, and computer science (CWCS-Data, 2021; Johnson et al., 2021).

Collections of resources and community can be found on the web pages for the American Institute of Physics' National Task Force to Elevate African American Representation in Undergraduate Physics & Astronomy (TEAM-UP) Project, and the AAAS's STEMM Equity Achievement (SEA) Change (AAAS, 2021; AIP, 2022; TEAM-UP, 2020).

<u>Do a New Thing</u>. Physicists at all levels can take it upon themselves to learn more about power and injustice across our discipline and with our organizations. As discussed in the majority of the <u>Put Your Big Kid Pants On</u> section, as well as the ***Policing and Gatekeeping*** paper, individual efforts matter (Hodari et al., 2022b). Namely, we can all learn more about how our behaviours, policies, and praxis impact those on the losing side of oppression, particularly when these appear neutral on the surface.

We are not absolved of this duty because we "happen" to live and work in a relatively homogenous community. [See (DiAngelo & Menakem, 2020) for more on whether homogeneity of really accidental.] It is our responsibility to each other to create in ourselves and our organizations the capacity to ensure that *any* interested person that enters can thrive. In the words of scholar Angela Johnson, we need to become "good detectors, so that when the rare unexpected event floats through, we don't miss it" (Johnson, 2018).

A common complaint of STEM faculty trying to diversify their departments is the difficulty attracting white women and BIPOC people because their local area is not demographically diverse (Hodari, 2001). We suggest that while context matters, the attraction of a positive experience outweighs context. People do not clamour to attend the Iowa Writers Workshop or the Clarion Writers Workshop (particularly before they moved to San Diego) because the surrounding area is demographically diverse. These programs attract participants because of the quality of the experiences they offer, and good news travels.

### The Long and the Short of It

<u>Monday Morning</u>. At the National Science Foundation's Joint Annual Meeting 2012 on Broadening Participation Research, STEM diversity leader Kelly Mack facilitated a panel discussion on women of color in STEM. After the panellists presented, and answered clarifying questions, Dr Mack opened the discussion by asking the audience to consider, "What can you do on Monday morning to broaden participation in your department? Her question remains apt. From your experience of this white paper, how would you answer Dr Mack's question across all forms of oppression that are salient in your professional contexts? Consider what ideas, strategies, or practices can you implement immediately? And perhaps more importantly, why haven't you done any of these already?

<u>Class of 2031</u>. Looking further ahead, imagine you are working with colleagues to contribute to Snowmass 2031. You likely have clear ideas about how you would like your science to progress in the intervening ten years. For the topics discussed in this white paper, what would you like to see? How will you contribute to improving the organizational context in which you work? What challenges might we face as we engage Snowmass 2031 because of what we do now?

$\mathcal{F}$rom the authors' perspective, it is far past time for us to stop pushing talented people out of physics because of who they are; or more accurately, <u>because of who we are</u> (Glaude, 2019).

### Acknowledgements
The authors acknowledge the Heising-Simons Foundation Grant #2020-2374 which supported this work. We also thank all the non-author organizers for the 2021 A Rainbow of Dark Sectors conference: Regina Caputo, Djuna Croon, Nausheen Shah, and Tien-Tien Yu. We additionally thank Risa Wechsler for her organizational support.